\documentclass[11pt]{article}
\usepackage[left=1.5cm, right=1.8cm]{geometry}
\usepackage{amsmath}
\usepackage{amssymb}
\usepackage{graphicx}
\usepackage{float}
\usepackage{authblk}

\usepackage{multirow}
\usepackage{epsfig}

\def\RR{{\mathfrak R}}

\newcommand{\Dslash}{D\mkern-11.5mu/\,} 
\newcommand{\delslash}{{\partial\mkern-9mu/}}

\def\CC{{\cal C}}
\def\Dbarslash{\,\,{\raise.15ex\hbox{/}\mkern-12mu {\bar D}}}
\def\Dslash{\,\,{\raise.15ex\hbox{/}\mkern-12mu D}}
\def\delslash{\,\,{\raise.15ex\hbox{/}\mkern-9mu \partial}}
\def\delbarslash{\,\,{\raise.15ex\hbox{/}\mkern-9mu {\bar\partial}}}

\newcommand{\SP}[1]{\begin{equation}\begin{split} #1
\end{split}\end{equation}}

\newcommand{\beq}{\begin{equation}}
\newcommand{\eeq}{\end{equation}}
\newcommand{\beqs}{\begin{eqnarray}}
\newcommand{\eeqs}{\end{eqnarray}}
\newcommand{\lsim}{\mathrel{\raisebox{-
.6ex}{$\stackrel{\textstyle<}{\sim}$}}}
\newcommand{\gsim}{\mathrel{\raisebox{-
.6ex}{$\stackrel{\textstyle>}{\sim}$}}}

\def\Dslash{\raisebox{1pt}{$\slash$} \hspace{-9pt} D}
\def\hbar{\hspace{0pt}\raisebox{1pt}{$-$} \hspace{-7pt} h}

\def\di{\mbox{d}}
\def\r{\rho}

\newcommand{\be}{\begin{equation}}
\newcommand{\ee}{\end{equation}}
\newcommand{\bea}{\begin{eqnarray}}
\newcommand{\eea}{\end{eqnarray}}

\def\lbldef#1#2{\expandafter\gdef\csname #1\endcsname {#2}}

\def\href#1#2{#2}

\newcommand{\ber}{\begin{eqnarray}}
\newcommand{\eer}{\end{eqnarray}}

\newcommand{\beqar}{\begin{eqnarray}}

\newcommand{\eeqar}{\end{eqnarray}}

\newcommand{\dsl}
  {\kern.06em\hbox{\raise.15ex\hbox{$/$}\kern-.56em\hbox{$\partial$}}}

\newcommand{\eeqarr}{\end{eqnarray}}
\newcommand{\ZZ}{{\rm \kern 0.275em Z \kern -0.92em Z}\;}

\def\CC{{\mathchoice
{\rm C\mkern-8mu\vrule height1.45ex depth-.05ex
width.05em\mkern9mu\kern-.05em}
{\rm C\mkern-8mu\vrule height1.45ex depth-.05ex
width.05em\mkern9mu\kern-.05em}
{\rm C\mkern-8mu\vrule height1ex depth-.07ex
width.035em\mkern9mu\kern-.035em}
{\rm C\mkern-8mu\vrule height.65ex depth-.1ex
width.025em\mkern8mu\kern-.025em}}}

\def\RR{{\rm I\kern-1.6pt {\rm R}}}

\def\ZZ{{\rm Z}\kern-3.8pt {\rm Z} \kern2pt}
\def\IB{\relax{\rm I\kern-.18em B}}
\def\ID{\relax{\rm I\kern-.18em D}}
\def\II{\relax{\rm I\kern-.18em I}}
\def\IP{\relax{\rm I\kern-.18em P}}

\newcommand{\bear}{\begin{eqnarray}}
\newcommand{\eear}{\end{eqnarray}}

\def\r{\rho}                                     

\def\6{\partial}

\def\bea{\begin{eqnarray}}
\def\eea{\end{eqnarray}}

\def\beqx{\begin{displaymath}}
\def\eeqx{\end{displaymath}}

\newcommand{\bmat}{\left(\begin{array}}
\newcommand{\emat}{\end{array}\right)}

\def\r{\rho}

\def\bo{{\raise-.3ex\hbox{\large$\Box$}}}               
\def\face{{\raise.2ex\hbox{$\displaystyle \bigodot$}\mskip-2.2mu \llap {$\ddot
        \smile$}}}                                   
\def\>{\rangle}                                      
\def\<{\langle}                                      

\def\slash#1{\rlap{\hbox{$\mskip 1 mu /$}}#1}        
\def\leftrightarrowfill{$\mathsurround=0pt \mathord\leftarrow \mkern-6mu
        \cleaders\hbox{$\mkern-2mu \mathord- \mkern-2mu$}\hfill
        \mkern-6mu \mathord\rightarrow$}        
\def\dvec#1{\vbox{\ialign{##\crcr
        \leftrightarrowfill\crcr\noalign{\kern-1pt\nointerlineskip}
        $\hfil\displaystyle{#1}\hfil$\crcr}}}           

\def\-{\hphantom{-}}

\begin{document}

\title{A composite light scalar, electro-weak symmetry breaking \\ and the recent LHC searches.}
\author{Daniel Elander}
\affil{Department of Theoretical Physics, Tata Institute of Fundamental Research, \\
Homi Bhabha Road, Mumbai 400 005, India}
\author{Maurizio Piai}
\affil{Swansea University, College of Science, \\
Singleton Park, Swansea, Wales, UK.}
\date{\today}
\maketitle

\abstract{We construct a model in which electro-weak symmetry breaking 
is induced by a strongly coupled sector, which is described in terms
of a five-dimensional model in the spirit of the bottom-up approach to holography.
We compute the precision electro-weak parameters, and identify  regions of parameter space
allowed by indirect tests. We compute the spectrum of scalar and vector resonances,
which contains a set of parametrically light states that can be identified with the
electroweak gauge bosons and a light dilaton. There is then a little desert, up to 2-3 TeV,
where towers of resonances of the vector, axial-vector and scalar particles appear.}

\maketitle

\newpage

\tableofcontents

\vspace{0.5cm}

\section{Introduction.}

The biggest open problem in high energy physics is to understand the origin of electro-weak symmetry breaking (EWSB) and hence of mass generation in the Standard Model (SM). 
Some new physics (particles and interactions) must be present at the electro-weak scale.
The main physics goal of the Large Hadron Collider (LHC)  is to collect as much experimental information as
possible about electro-weak scale new physics, in order to address this problem.

From a theoretical point of view, the most fundamental question is whether the new interactions are weakly coupled, and hence EWSB originates from the condensation of  elementary scalar fields 
(as is the case in the minimal version of the SM), or
whether the new interaction is strong enough to trigger the formation of non-trivial composite condensates,
as is the case in QCD regarding chiral symmetry breaking (as in Technicolor and its variations~\cite{TC,WTC,ETC,reviews}).
From the experimental point of view, the most urgent question is whether a light scalar particle
is present in the spectrum, the couplings of which are related to the masses of the other particles.
In the (weakly coupled) minimal version of the SM, this  is the Higgs particle.
In certain (strongly coupled) models of dynamical electro-weak symmetry breaking,
a particle with similar properties may be present as a consequence of the spontaneous breaking of 
(approximate) scale invariance: the dilaton~\cite{dilaton,dilatonpheno,dilaton2,dilaton5D}.

The main purpose of this paper is to provide a simple example of a model
of the second class: there is no elementary scalar field, the fundamental interactions
are strongly-coupled, and EWSB is due to the formation of some non-trivial condensate.
However, because the model is approximately scale invariant, the condensate breaks spontaneously 
such symmetry, and as a result the spectrum contains a light dilaton.

Before the LHC program started, indirect searches 
for new physics have been carried out for quite some time, after the discovery of the $W$ and $Z$
gauge bosons. The surprising result of these searches is that, in spite of the fact
that new interactions and new particles must be responsible for the weak scale $v\simeq 246$ GeV,
none of these
 particles produces a sizable contribution to precision physics observables~\cite{Peskin,Barbieri},
with the exception of a, possibly light, Higgs particle.
This suggests the existence of a little desert in the spectrum, with a light Higgs followed by
a  gap, and possible new particles appearing only with masses in the $2-5$ TeV range.
The dynamical generation and stabilization of such a gap 
is often referred to as the {\it little hierarchy} problem.

In order for a strongly-coupled model 
to be viable, we must be able to compute the precision parameters, the mass of the light
states such as the $W$, $Z$ and the possible light scalars,
but also the mass spectrum of the heavy resonances,
and show that all of these are in agreement with experimental bounds.
This is a quite challenging program.

Gauge/gravity dualities~\cite{AdSCFT,Itzhaki:1998dd,reviewAdSCFT} 
allow, in principle, for such calculations to be done.
The idea is that one constructs a consistent background of an extra-dimension theory of gravity
(string theory), and use it to compute quantities, such as correlation functions,
in the dual four-dimensional gauge theory
without ever referring to the microscopic Lagrangian of the latter.
A difficulty of this approach is that it is still hard to study the dual of confining gauge
theories. Notable examples exist, based on the deformations of the sphere~\cite{PS,gppz},
of the conifold (a sample of the vast literature on the subject includes for instance~\cite{CO,KW,KT,KS,MN,PT,conifolds,consistentconifold}) 
and of the Witten-Sakai-Sugimoto system~\cite{SS}.
But the systematics of all of these models is difficult to understand, 
hard to generalize, and calculations are technically  challenging.

A simpler approach to holography consists of the so called bottom-up phenomenological models.
Rather than relying on a fully consistent stringy construction, these models 
assume a set of simplifying assumptions for a five-dimensional system, and then proceed to
perform the same kind of calculations one would do in a consistent gravity background.
The advantage is obvious: constructing such models is much easier, and one can easily 
study whole classes of models by just dialing the free parameters.
The disadvantage is also clear: not being derived from a fundamental theory of gravity,
these models are somewhat arbitrary, and contain many more free parameters.
This is  reminiscent of what happens in effective theories, where one
deals with large number of free parameters, all of which have the same fundamental origin,
and hence would be correlated, if one could compute them from first principles.

Nevertheless, these models, because of their simplicity, are a very powerful tool,
because they allow to ask how general certain results are, without having to rely on 
specific points of a parameter space that is otherwise hard to explore.
The literature on the topic is rather extensive (for examples related to technicolor see~\cite{AdSTC}).
We are going to build one such a model. The main differences
with respect to what done in other papers is that the background is 
not AdS, and that we use a construction which is simple enough to allow to study
at the same time both the spin-1 and spin-0 sectors.
  
The five-dimensional gravity theory is coupled to a scalar field, for which we assume
a peculiarly simple (super)potential. We solve the coupled equations and find that the metric
is asymptotically AdS both in the far UV and in the deep-IR, resembling the dual 
description of the field theory flow between two fixed points (see for example~\cite{PW,LS}).
We will see that the results and the numerology of such a setup are indeed not 
dissimilar from the rigorous results related to the PW flow (for example, our spectrum resembles
qualitatively the one obtained in~\cite{KMPP}).
Because confinement is the hard thing to model, we introduce a hard cutoff in the IR,
which introduces a scale $\Lambda_0$ in the theory.
With abuse of language, we will call this the confinement scale, although in a rigorous sense this
is rather just an IR regulator. 

We then proceed to introduce the electro-weak group and electro-weak symmetry breaking. 
In the top-down approach, this can be done by introducing probe-branes, 
and showing that the curvature yields a $U$-shape embedding for the brane which can be interpreted
 in terms of the spontaneous breaking $U(1)\times U(1)\rightarrow U(1)$ 
 (along the lines of what proposed in ~\cite{SS}).
 We mimic this procedure, by allowing the gauge bosons of $U(1)_L\times U(1)_R$
 to propagate in the bulk, and imposing different boundary conditions in the IR
 for the axial and vector combination of the two fields, hence inducing the desired breaking pattern.
 Notice that the fact of not gauging a $SU(2)_L$ means that there are no $W$ bosons in the spectrum.
 However, it would be trivial to extend the construction, at the only price of complicating the algebra.
 For our purposes, this is going to be satisfactory.
 
 We then compute the spectrum of the four-dimensional spin-1 states. 
 After holographic renormalization~\cite{HR},  it consists of two light states 
 (that are identified with the massless photon and the massive $Z$ boson),
 and two towers of heavy composite spin-1 states, the techni-$\r$ (vectorial) and techni-$a_1$ (axial) mesons,
 with masses controlled by the scale $\Lambda_0$.
We also compute the oblique  precision parameters, compare to the experimental bounds, and 
hence fix some parameter in the model.

We then move on to study the scalar fluctuations of the background.
This non-trivial task is made peculiarly simple by the fact that the formalism for doing so has been developed in detail in the literature~\cite{dilaton5D,BHM,EP}, and by the fact that in the presence of one scalar only, with dynamics controlled by a superpotential, the equations become manageable.
We compute the spectrum of perturbations, and observe that it always consists 
of a tower of heavy states, with mass controlled by $\Lambda_0$,  supplemented
by an additional, parametrically light state (the dilaton).

We  discuss in some detail under what circumstances such a dilaton is light.
Its presence is due to the fact that the IR cutoff effectively amounts to the spontaneous breaking
of scale invariance. The fact that the background has a non-trivial profile, interpolating between two AdS
spaces with different curvature, can be interpreted (depending on the parameters) 
in terms of the presence of explicit breaking of scale invariance, which induces a mass.
However, notice that because of the rough way in which we model confinement,
our result is to be understood as a lower bound: other sources of explicit breaking 
will be present in a complete confining theory. It is remarkable that light states have been seen 
within the top-down approach (for example in~\cite{ENP}), but as we said, the systematics of this is 
not well understood.

We conclude the paper with  numerical examples.
We show that it is remarkably easy to produce a model
in which the precision parameters are within the experimental bounds, 
in which a light scalar has a mass of the order of the $Z$ boson, and in which 
a little desert follows, with heavy resonances appearing with masses above a few TeV.

The main message of the paper is that discovering (or not) a light scalar, with 
couplings similar to the Higgs particle
of the Standard Model, does not imply that EWSB is  (or is not) due to a new weakly coupled interaction. 
We want  to show that it is actually easy to construct strongly coupled models that mimic such a scenario.
In order to distinguish the two scenarios, the LHC experiments 
should explore in detail the whole  mass region up to 2-4 TeV
range, at high luminosity, and see whether other new resonances are there.
A task  within reach, provided the LHC increases its operational energy and
luminosity, all of which is according to the future plan of the LHC program.

\section{General set-up.}

The model we study consists of a five dimensional theory, 
in which gravity, a scalar field $\Phi$, and the $U(1)\times U(1)$ gauge bosons ($L_M$ and $R_M$)
propagate in a background geometry.
We use the following conventions, as in~\cite{EP}, and follow the same formalism. 
Capital roman indexes $M=0,1,2,3,5$ are five-dimensional space-time indexes,
while greek indexes $\mu=0,1,2,3$ are restricted to the 4-dimensional Minkowski slices of the space.
In this way, we label the space-time coordinates as $x^M=(x^{\mu},r)$, with $r$ the radial (fifth) direction. 
We write the five-dimensional metric $g_{MN}$ with signature $\{-++++\}$.
We assume that the background metric satisfies the ansatz
\beqs
\label{Eq:bgmetric}
\di s^2&=&e^{2A}\eta_{\mu\nu}\di x^{\mu}\di x^{\nu}+\di r^2\,,
\eeqs
with $A=A(r)$. 
Finally, we assume that the radial direction is compact, with $r_1<r<r_2$.
The boundaries are interpreted as an IR cutoff $r_1$ (which models in a very crude way
the confinement scale of the putative four-dimensional dual field theory),
and a UV cutoff $r_2$ (which is interpreted as a regulator, and will be removed 
at the end of the calculations via holographic renormalization).

The geometry is determined by the dynamics of the scalar field $\Phi$,
which represents a simple example of five-dimensioanl sigma-model coupled to gravity.
We write the action as
\beqs
{\cal S}&\equiv&\int\di^4x\di  r \Bigg\{
\sqrt{-g}
\Theta
\left[\frac{1}{4}R+{\cal L}_5(\Phi,\partial_M\Phi,{g}_{MN}) \right] 
\nonumber\\
&&\left.
+\sqrt{-\tilde{g}}\delta( r- r_1) \left[c_KK+{\cal L}_1(\Phi,\partial_{\mu}\Phi,\tilde{g}_{MN}) \right] \right.\nonumber\\
&&
- \sqrt{-\tilde{g}} \delta( r- r_2)\left[c_KK+{\cal L}_2(\Phi,\partial_{\mu}\Phi,\tilde{g}_{MN}) \right] 
\Bigg\}\,,
\label{Eq:action}
\eeqs
where  $R$ is the Ricci scalar, where
$K$ is the extrinsic curvature, where the coupling $c_K=-1/2$ is fixed by consistency and where
${\cal L}_i$ are the sigma-model actions.
The step function is defined by $\Theta\equiv\Theta( r- r_1)-\Theta( r- r_2)$.

We define the action of the matter fields in terms of the real scalar field $\Phi=\Phi(x^{\mu}, r)$ as
\beqs
\label{Eq:L5}
{\cal L}_5&\equiv&-\frac{1}{2}g^{MN}\partial_M\Phi\partial_N\Phi-V(\Phi)\,,\\
\label{Eq:L1}
{\cal L}_1&\equiv&-\lambda_{(1)}(\Phi)\,,\\
\label{Eq:L2}
{\cal L}_2&\equiv&-\lambda_{(2)}(\Phi)\,.
\eeqs
For simplicity, we take the boundary actions for the scalar to depend only on the field, and not on the
derivatives.

Assuming that there exists a superpotential $W$ such that the potential $V$
 can be rewritten as
\beqs
V&=&\frac{1}{2}\left({\partial_{\Phi} W}{}\right)^2\,-\,\frac{4}{3}W^2\,,
\eeqs
where $\partial_{\Phi}W\equiv\frac{\partial W}{\partial \Phi}$, 
then the system can be reduced to first-order equations
\beqs
\label{Eq:BPS1}
A^{\prime} &=& -\frac{2}{3} W\,,\\
\label{Eq:BPS2}
\bar{\Phi}^{\prime}&=&{\partial_{\Phi} W}\,,
\eeqs
in the sense that all solutions to Eqs.~(\ref{Eq:BPS1})-(\ref{Eq:BPS2}) are also solutions to the second-order equations derived from ${\cal S}$.\footnote{Given a five-dimensional gauged supergravity for which a superpotential $W$ is known, the first-order equations Eqs.~(\ref{Eq:BPS1})-(\ref{Eq:BPS2}) agree with the BPS equations one would obtain by studying the supersymmetry variations of the gravitino and dilatino. Hence, solving Eqs.~(\ref{Eq:BPS1})-(\ref{Eq:BPS2}) would amount to finding supersymmetric backgrounds. The converse is not true: the existence of $W$, and of solution to Eqs.~(\ref{Eq:BPS1})-(\ref{Eq:BPS2}), does not imply that the model is supersymmetric. In the present context, the fact that we study a model defined by a given choice of $W$ is to be understood as just a technical tool, yielding a simple solution-generating technique. Notice in particular that the model we study does not contain any fermions, and that the fluctuations of the scalars are described by second-order equations. Our study does not assume nor imply that the strongly-coupled dual field theory be supersymmetric, in particular the existence of a light composite scalar is not due to supersymmetry.}

We call $\Phi_1=\Phi(r_1)$ and $\Phi_2=\Phi(r_2)$ the boundary values of the 
scalar, which we use to define the boundary conditions. 
Accordingly, the general form of the localized potentials must be
\beqs
\lambda_{(1)}&=&W(\Phi_1)\,+\,\partial_{\Phi}W(\Phi_1)({\Phi}-\Phi_1)\,+\,\frac{1}{2}\partial_{\Phi}^2\lambda_{(1)}({\Phi}-\Phi_1)^2\,,\\
\lambda_{(2)}&=&W(\Phi_2)\,+\,\partial_{\Phi}W(\Phi_2)({\Phi}-\Phi_2)\,+\,\frac{1}{2}\partial_{\Phi}^2\lambda_{(2)}({\Phi}-\Phi_2)^2\,.
\eeqs
In this way, the whole system can be described in terms of 
first order equations which satisfy automatically the second-order equations derived by
variational principle from the five-dimensional action.

For reasons that will become clear later, we find it convenient to define:
\beqs
N&\equiv&\partial_{\Phi}^2W - \frac{(\partial_{\Phi}W)^2}{W} \,,\\
\partial_{\Phi}^2\lambda_{(1)}&\equiv&\left.\partial_{\Phi}^2W\right|_{r_1}\,+\,\left(m^{2}_{1}\right)\,,\\
\partial_{\Phi}^2\lambda_{(2)}&\equiv&\left.\partial_{\Phi}^2W\right|_{r_2}\,-\,\left(m^{2}_{2}\right)\,,
\eeqs
where $m_i^2$ are arbitrary mass terms that we will (conservatively) choose to 
diverge $m_i^2\rightarrow +\infty$.

The gauge bosons are introduced as probes, which are not back-reacted on the metric,
via the following action:
\beqs
{\cal S}_{5}&\propto&-\frac{1}{4}\int\di^4x\int_{r_1}^{r_2} \di r \left[\frac{}{}\left(\frac{}{}a (r)+b(r)D\delta(r-r_2)\right)\eta^{\mu\nu}\eta^{\gamma\delta}F_{\mu\gamma}F_{\nu\delta}+2b (r)\eta^{\mu\nu}F_{r\mu}F_{r\nu}\right]\,,
\eeqs
where $F$ stands for the field-strength of both the $V$ and $A$ fields, which are defined by
\beqs
L&=&\cos\theta_W A + \sin\theta_W V\,,\\
R&=&-\sin\theta_W A + \cos\theta_W V\,.
\eeqs
The gauge couplings $g$ and $g^{\prime}$ of the $U(1)_L$ and $U(1)_R$, respectively,
determine the mixing angle $\tan\theta_W=\frac{g^{\prime}}{g}$ that relates the $(L,R)$
basis and the $(V,A)$ basis.

After Fourier-transforming in the four Minkowski directions (and writing the four-momentum $q^{\mu}$
with the opposite convention with signature $(+,-,-,-)$ for later convenience),
we can  rewrite $A^{\mu}(q,r)\rightarrow A^{\mu}(q)v(q,r)$, in such a way that
all the information about the bulk theory be contained in the functions $v(q,r)$.
The two kinds of fields have the same equations of motion, given by
\beqs
\partial_{r}\left(\frac{}{}b(r)\partial_{r}v(q,r)\right)&=&-q^2 a(r)v(q,r)\,.
\eeqs
Electro-weak symmetry breaking is induced by the fact that
the solutions $v_{A,V}$ must satisfy {\it different} IR boundary conditions:
\beqs
v_A(r_1)&=&0\,=\,\partial_{r}v_V(r_1)\,.
\eeqs

For the functions $a$ and $b$ we take the simplest  possible form,
which can be thought of as coming from allowing the spin-1 fields to propagate in the 
space with the geometry determined by the five-dimensional sigma-model background:
\beqs
a(\r)&=&\sqrt{-g}g^{xx}g^{xx}\,=\,1\,,\\
b(\r)&=&\sqrt{-g}g^{xx}g^{rr}\,=\,e^{2A}\,.
\eeqs

\section{The model.}

We borrow model C from~\cite{EP}, for which
\beqs
W&=&-\frac{3}{2}-\frac{\Delta}{2}\Phi^2+\frac{\Delta}{3\Phi_I}\Phi^3\,,
\eeqs
and we focus on the solution
\beqs
\bar{\Phi}&=&\frac{\Phi_I}{e^{\Delta(r-r_{\ast})}+1}\,,
\eeqs
which as expected is a 1-parameter family labelled by $r_{\ast}$.

In this case, evaluating on the solution, one has
\beqs
W&=&
\frac{1}{6} \left(-9 - \frac{
   \Delta (1 + 3 e^{\Delta (r - r_{\ast})} )\Phi_I^2}{(1 + e^{\Delta (r - r_{\ast})})^3}\right)\,,\\
   W^{\prime} &=&
   -\frac{\Delta e^{\Delta (r + r_{\ast})} \Phi_I}{(e^{\Delta r} + e^{\Delta r_{\ast}})^2}\,,\\
   A&=&\frac{1}{9} \left(9 r + 
   \Phi_I^2 \frac{e^{\Delta (r + r_{\ast})}}{(e^{\Delta r} + e^{\Delta r_{\ast}})^2} + \Phi_I^2\Delta r - 
   \Phi_I^2 \log[1 + e^{\Delta (r - r_{\ast})}]\right)\,\\
   &&\nonumber -\frac{1}{9} \Phi_I^2 \left(\frac{1}{2 \cosh (\Delta
  r_{\ast})+2}-\log \left(1+e^{-\Delta r_{\ast}}\right)\right)\,.
\eeqs
Notice that  we conventionally fix $r_1=0$  from now on, and 
we chose an integration constant in $A$ so that $A(0)=0$.
In this way, all the masses will be expressed in units of an overall scale $\Lambda_0$,
that will be fixed by the mass of the $Z$ boson.

Before we start to study the physics of the model we do some parameter counting.
The parameters in the model are: $r_1$, $r_2$, $\Phi_I$, $\Delta$, $m_{1,2}^2$, $r_{\ast}$, $D$ and
an overall constant in front of ${\cal S}_5$.
The constant in front of ${\cal S}_5$ is related to the effective five-dimensional 
gauge coupling, and we will later call it ${\cal N}$. This constant together with $D$ effectively
control the strength of the gauge couplings of the light degrees of freedom (photon and $Z$ boson),
and of the resonances.
Because we take the gauge couplings of the lightest states to be identified with the standard-model
couplings, there is only one free parameter, which is the ratio of these
 gauge couplings to the effective couplings of the resonances 
 (one can think of the latter as the equivalent to the $g_{\r\pi\pi}$ coupling of QCD).
We set $r_1=0$, hence setting all mass scales in terms of the same 
scale, which enters both into quantities related to confinement and to symmetry breaking.
$r_2$ is a UV-regulator, which will be taken to infinity at the end of the calculations, and hence has no 
physical meaning. The boundary masses $m_{1,2}^2$ do not enter in the equations determining the background, but only in the equations for the scalar fluctuations. We will take them to diverge, as anticipated,
as a conservative assumption (if we find a light state in the spectrum of scalar perturbations
in this case, it will in general be parametrically light also for any other choice of $m_{1,2}^2$~\cite{EP}).
Concluding: there are four parameters that are  physical and free: $\Phi_I$, which controls the departure
from AdS of the bulk geometry, $\Delta$ which can be interpreted in terms of the dimensionality of the dual operator inducing the departure from conformality, $r_{\ast}$ which controls the physical scale
at which the departure from conformality arises, and which is independent of the confinement scale,
and the effective coupling of the spin-1 resonances, which can be interpreted in terms of the 
large-$N$ expansion parameter of the dual field theory.

A comment about the model. The remarkable simplicity of the superpotential
clearly indicates that this is not a model that has been obtained from a complete string
theory with the familiar tools of consistent truncation. However, the background
that it yields has many features that are very similar to those obtained in 
more rigorous constructions (for example, see~\cite{PW}). The simplicity
of this model allows us to perform the actual calculations (almost) analytically,
and we expect that the main features we obtain could be derived from a large
class of string-motivated models.

\section{Electroweak symmetry breaking.}

What we are constructing is the low energy description of a simplified
model of  dynamical electroweak symmetry breaking, in which  as anticipated
we simplify the standard model by replacing the $SU(2)_L\times U(1)_Y\rightarrow U(1)_{\rm e.m.}$
with $U(1)_L\times U(1)_R\rightarrow U(1)_V$, with gauge couplings $g$ 
and $g^{\prime}$ for the $L$ and $R$ gauge groups, respectively.
The Lagrangian of the two light gauge bosons,
focusing on the transverse vacuum polarizations,
can be written in general as
\beqs
{\cal L}_4 &=&-\frac{1}{2}P^{\mu\nu}A_{\mu}^{i}(q)A_{\nu}^{j}(-q)\pi(q^2)_{ij}\,,
\eeqs
where $i=L,R$, where $P^{\mu\nu}=\eta^{\mu\nu}-q^{\mu}q^{\nu}/q^2$
and where the $\pi^{i,j}$ are (possibly complicated) functions of the four-momentum.
Such functions are going to be non-anaytical in general. However, under reasonable assumptions,
they will be regular at the origin, and can be expanded as
\beqs
\pi(q^2)&=&\pi(0)+q^2 \pi^{\prime}(0)+\frac{1}{2}(q^2)^2\pi^{\prime\prime}(0)+\cdots\,.
\eeqs

The precision electro-weak parameters are defined by the following relations~\cite{Barbieri}:
\beqs
\hat{S}&=&\frac{g}{g^{\prime}}\pi^{\prime}_{LR}(0)\,,\\
X&=&\frac{1}{2}M_W^2\pi^{\prime\prime}_{LR}(0)\,,\\
W&=&\frac{1}{2}M_W^2\pi^{\prime\prime}_{LL}(0)\,,\\
Y&=&\frac{1}{2}M_W^2\pi^{\prime\prime}_{RR}(0)\,,
\eeqs
where $M_Z$ is the mass of the lightest axial state, while we just define
$M_W^2\equiv\cos^2\theta_W M_Z^2$.
Notice that because we restrict our attention to the neutral gauge bosons, 
the custodial-symmetry breaking parameters $\hat{T}$ and $\hat{U}$ do not exist.\footnote{In order to avoid generating a significant contribution to electroweak precision parameters related to custodial symmetry breaking, one might have to extend the bulk symmetry to $SU(2) \times SU(2)$. In doing so one would ensure that the $\hat{T}$ and $\hat{U}$ parameters are negligibly small, independently of the bulk dynamics. See for example~\cite{Round}, and references therein, for a general discussion of this and related issues, which we do not address in our study.}
In these definitions, one assumes that the fields are normalized 
so that $\pi_{LL}^{\prime}(0)=1=\pi_{RR}^{\prime}(0)$.
By making use of  the rotations from the $(L,R)$ basis to the
vector axial-vector basis $(V,A)$,
 we can rewrite
\beqs
\hat{S}&=&\cos^2\theta_W \left(\frac{}{}\pi_V^{\prime}(0)-\pi_A^{\prime}(0)\right)\,,\\
X&=&\cos^2\theta_W\frac{M_Z^2}{2} \sin\theta_W\cos\theta_W\left(\frac{}{}\pi_V^{\prime\prime}(0)-\pi_A^{\prime\prime}(0)\right)\,,\\
W&=&\cos^2\theta_W\frac{M_Z^2}{2} \left(\frac{}{}\sin^2\theta_W\pi_V^{\prime\prime}(0)+\cos^2\theta_W\pi_A^{\prime\prime}(0)\right)\,,\\
Y&=&\cos^2\theta_W\frac{M_Z^2}{2}  \left(\frac{}{}\cos^2\theta_W\pi_V^{\prime\prime}(0)+\sin^2\theta_W\pi_A^{\prime\prime}(0)\right)\,.
\eeqs
All of this means that having the functions $\pi_{V,A}$ up to the second derivative in 
$q^2$ is enough to yield all the precision parameters.

The fit of the experimental data from precision electroweak physics 
indicates that all of  these four precision parameters 
must be at most of the order of $3\times 10^{-3}$~\cite{Barbieri}.
The precise results depend on what confidence level we use  and on what value 
of the reference mass for the Higgs particle we adopt, and the various bounds are correlated. 
For the present purposes, 
this very rough  bound is sufficient.
One can see that (at least approximately), provided the precision parameters
are all small (which in turns means that $\pi^{\prime}_A\simeq \pi_V^{\prime}=1$)
the symmetry-breaking scale
can be measured by the mass of the lightest axial mode:
\beqs
M_Z^2&\simeq&-\frac{\pi_A(0)}{\pi_A^{\prime}(0)}\,\simeq\,-{\pi_A(0)}\,.
\eeqs

Before focusing on the specific model of this paper, let us remind  the reader about
how to perform the calculation of the precision parameters in the general case 
of a five-dimensional model of this type.
For the vectorial bosons, it is convenient to define the following:
\beqs
\partial_{r} v(q,r)&\equiv&\gamma(q,r)v(q,r)\,,
\eeqs
and replace in the bulk equations, which yield
\beqs
\partial_{r}(b\gamma)+b\gamma^2+q^2 a &=&0\,,
\eeqs
where we simplified the notation by dropping obvious terms.
This first-order equation contains the same amount of information as the original 
second-order equation, but it has the advantage that the overall normalization
of the state $v$ drops.
The functions $\gamma$ must satisfy the boundary condition:
\beqs
\gamma\left(q,r_1\right)&=&0\,.
\eeqs

For the axial part, it is convenient to define the inverse of the function $\gamma$, via
\beqs
v(q,r)&\equiv&X(q,r)\partial_{r}v(q,r)\,,
\eeqs
which obeys the (same) equation
\beqs
-b^2\partial_{r}\left(\frac{X}{b}\right) + b + q^2 a X^2&=&0\,, 
\eeqs
subject to the (different) boundary condition
\beqs
X\left(q,r_1\right)&=&0\,.
\eeqs

One hence must find solutions to the equations for $\gamma$ and $X$, subject to 
the boundary conditions in the deep-IR.
Finally, the physical spectrum and the two-point correlation-functions can be studied 
from evaluating the solutions at the UV-boundary $r_2$, 
and then trying to take the limit $r_2\rightarrow +\infty$, if possible.
Before doing so, some more about solving the actual equations.

For the calculation of the precision parameters, we need only the first few terms
of the expansion of $\gamma$ and $X$ in $q^2$.
We define the low-energy expansions
\beqs
\gamma(q,r)&\equiv&\gamma_0(\r)+q^2\gamma_1(\r)+\frac{1}{2}(q^2)^2 \gamma_2(\r)+\cdots\,,\\
X(q,\r)&\equiv&X_0(\r)+q^2X_1(\r)+\frac{1}{2}(q^2)^2 X_2(\r)+\cdots\,,
\eeqs
and expand accordingly the equations of motion which yield
\beqs
\partial_{r}(b \gamma_0) + b \gamma_0^2&=&0\,,\\
\partial_{r}(b \gamma_1) +2b \gamma_0\gamma_1 + a &=&0\,,\\
\partial_{r}(b \gamma_2)+2b (\gamma_0\gamma_2+\gamma_1^2)&=&0\,,\\
&\cdots&\nonumber\,,\\
 -b^2\partial_{r}\left(\frac{X_0}{b}\right) + b &=&0\,,\\
 -b^2\partial_{r}\left(\frac{X_1}{b}\right)  +  a X_0^2&=&0\,,\\
 -b^2\partial_{r}\left(\frac{X_2}{b}\right)+4 a X_0X_1&=&0\,,\\
 &\cdots&\,,\nonumber
\eeqs
subject to the boundary conditions
\beqs
\gamma_i\left(r_1\right)&=&0\,,\\
X_i\left(r_1\right)&=&0\,,
\eeqs
which ensure that the IR boundary conditions are satisfied for every $q$.
A further simplification occurs thanks to the boundary conditions, which can be satisfied only provided 
$\gamma_0=0$ identically for any $r$. This reflects the physical fact that $\gamma$
is going to be related to the inverse of the propagator of the gauge boson of an unbroken $U(1)$,
and hence must contain a massless mode.

At least formally, the equations can be integrated, to yield:
\beqs
\gamma_0(r)&=&0\,,\\
\gamma_1(r)&=&-\frac{1}{b(r)}\int_{r_1}^{r}\di \omega a(\omega)\,,\\
\gamma_2(r)&=&-\frac{1}{b(r)}\int_{r_1}^{r}\di \omega 2b(\omega)\gamma_1(\omega)^2\,,\\
&\cdots&\,\nonumber\\
X_0(r)&=&b(\r)\,\int_{r_1}^r\frac{d \omega}{b(\omega)}\,,\\
X_1(r)&=&b(\r)\,\int_{r_1}^r\frac{a(\omega)X_0(\omega)^2}{b(\omega)^2}\,,\\
X_2(r)&=&b(\r)\,\int_{r_1}^r\frac{4a(\omega)X_0(\omega)X_1(\omega)}{b(\omega)^2}\,,\\
&\cdots&\,.\nonumber
\eeqs

We can now evaluate the effective five-dimensional action on-shell.
Doing so, because we are satisfying the bulk equations and the IR boundary conditions, 
leaves only a boundary term localized at the UV boundary $r_2$.
To this, we must add localized boundary actions, which depend only on the four-dimensional fields $A_{\mu}$ and $V_{\mu}$.
The results are
\beqs
\pi_V(0)&=&0\,,\\
\pi_A(0)&=&{\cal N}(r_2)\left(\frac{}{}C(r_2)b(r_2)+ b(r_2)\frac{1}{X_0(r_2)}\frac{}{}\right)\,,\\
\pi_V^{\prime}(0)&=&{\cal N}(r_2)\left(\frac{}{}D(r_2)b(r_2)+b(r_2)\gamma_1(r_2)\frac{}{}\right)\,,\\
\pi_A^{\prime}(0)&=&{\cal N}(r_2)\left(\frac{}{}D(r_2)b(r_2)
-b(\r_2)\frac{X_1(r_2)}{X_0(r_2)^2}\frac{}{}\right)\,,\\
\pi_V^{\prime\prime}(0)&=&{\cal N}(r_2)\left(\frac{}{}E(r_2)b(r_2)+b(r_2)\gamma_2(r_2)\frac{}{}\right)\,,\\
\pi_A^{\prime\prime}(0)&=&{\cal N}(r_2)\left(\frac{}{}E(r_2)b(r_2)+b(r_2)\left(\frac{2X_1(r_2)^2}{X_0(r_2)^3}
-\frac{X_2(r_2)}{X_0(r_2)^2}\right)\frac{}{}\right)\,,
\eeqs
where ${\cal N}$ is a normalization that is related to the wave-function normalization of the four-dimensional 
fields, and that is fixed by requiring that the photon be canonically normalized, while $C$, $D$ and $E$ are 
counterterms. Notice however that the counterterm $C$ naively breaks
 gauge-invariance, and hence must come (if non-vanishing) from a localized (elementary) Higgs field. 
The counterterm $D$ is always present,
and effectively controls the strength of the (weak) gauging of the global $U(1)$s.
The counterterm $E$ comes from higher-order operators of the form $(\partial F)^2$,
which would signal the fact that the theory is UV-incomplete.

Let us now move back to our example.
In our case $C=E=0$, $r_1=0$, $a=1$ and $b=e^{2A}$.
With all of this in place, one finds that 
\beqs
b(r_2)\gamma_1(r_2)&=&-\int_0^r\di \omega\,=\,-r_2\,,
\eeqs
which signals the presence of a (logarithmic) divergence in the kinetic term
of the vector fields. This is expected in full generality, and is connected with
the running of the (weak) four-dimensional couplings because
the elementary fields of the strongly-coupled dual theory are charged.
This allows us to choose the parameter $D$ to be
\beqs
D b(r_2)&=&r_2-\frac{1}{\varepsilon^2}\,,
\eeqs
effectively trading the divergence for a free parameter.
With the additional choice of normalization
\beqs
{\cal N}&=&-\varepsilon^2\,,
\eeqs
one recovers
\beqs
\pi^{\prime}_V&=&1\,.
\eeqs

Hence, for generic values of the parameters of the model, we have
\beqs
\pi_V(0)&=&0\,,\\
\pi_A(0)&=&-\varepsilon^2b(r_2)\left(\frac{}{} \frac{1}{X_0(r_2)}\frac{}{}\right)\,=\,
\frac{-\varepsilon^2}{\int_0^{r_2} e^{-2A(\omega)}\di \omega}\,,\\
\pi_V^{\prime}(0)&=&1\,,\\
\pi_A^{\prime}(0)&=&-\varepsilon^2\left(\frac{}{}r_2-\frac{1}{\varepsilon^2}
-\frac{b(r_2)X_1(r_2)}{X_0(r_2)^2}\frac{}{}\right)\,,\\
\pi_V^{\prime\prime}(0)&=&-\varepsilon^2b(r_2)\left(\frac{}{}\gamma_2(r_2)\frac{}{}\right)\,,\\
\pi_A^{\prime\prime}(0)&=&-\varepsilon^2b(r_2)\left(\frac{}{}\frac{2X_1(r_2)^2}{X_0(r_2)^3}
-\frac{X_2(r_2)}{X_0(r_2)^2}\frac{}{}\right)\,.
\eeqs
Notice in particular that because A is approximately linear at large-$r$,
 the integrals now all converge, and the 
limit $r_2\rightarrow +\infty$ can be taken.

Let us do all the calculations explicitly.
This can be done numerically, or with the following approximations 
put in place. It is  reasonable (though not necessary)
 to assume that $\Delta$ lies in the range  $1<\Delta<2$.
The holographic interpretation of this is that the bulk is not AdS because 
 the dual theory is deformed by the insertion of an  operator of dimension $4-\Delta$
 (for a concrete example, see again~\cite{PW}).
It is also reasonable to assume that $\Phi_I$ be at most ${\cal O}(1)$.
If so, the warp factor $A$ can be approximated very well by
\beqs
A&\simeq&
r \left(\frac{\Delta \Phi_I^2}{9}+1\right) \Theta
   (r_{\ast}-r)\\ && \nonumber+\Theta (r-r_{\ast}) \left(\frac{1}{9} \Delta
   r_{\ast} \Phi_I^2-\frac{1}{9} \left(\frac{1}{2 \cosh
   (\Delta r_{\ast})+2}-\log \left(1+e^{-\Delta
   r_{\ast}}\right)\right) \Phi_I^2+r\right)\,,\\
   &\equiv& w_1 r \Theta  (r_{\ast}-r) + \Theta  (r-r_{\ast})\left(r+w_2\right)\,,
\eeqs
where the constants $w_1$ and $w_2$ have a very different meaning (and dimensionality).
Notice that this is a very rough approximation, especially for small $\Delta$ and for $r\simeq r_{\ast}$.

Then, one finds that
\beqs
X_0(r)&=&e^{2A(r)}\int_0^r\di \omega e^{-2A(\omega)}
\nonumber\\
&\simeq&\frac{e^{2A(r)}}{2}\left[e^{-2(r_{\ast}+w_2)}-e^{-2(r+w_2)}+\frac{1-e^{-2w_1r_{\ast}}}{w_1}\right]\,,
\eeqs
provided $r>r_{\ast}>0$, while
\beqs
X_0(r)&=&e^{2A(r)}\int_0^r\di \omega e^{-2A(\omega)}\,\simeq\,
\frac{e^{2r}}{2}\left(1-e^{-2r}\right)\,
\eeqs
when $r_{\ast}\ll 0$.

From these, by taking $r_2\rightarrow \infty$, we can read
\beqs
M_Z^2&\simeq&-\pi_A(0)\,\simeq\,\frac{2\varepsilon^2 w_1}{1-e^{-2w_1r_{\ast}}+w_1e^{-2(r_{\ast}+w_2)}}\,,\\
&\simeq & 2\varepsilon^2 w_1\,,
\eeqs
for $r_{\ast}\gg0$ and 
\beqs
M^2_Z&\simeq& 2\varepsilon^2\,,
\eeqs
when $r_{\ast}\ll0$.
This is also the limit $\Phi_I\rightarrow 0$, in which  
the space becomes AdS with unit curvature, and $M_Z^2\simeq 2\varepsilon^2$.

The physical value of the $Z$ mass is $M_Z^{exp}\simeq 91$ GeV.
Equating the two, we find that $M_Z^{exp}=M_Z\Lambda_0$, where $\Lambda_0$
is the confinement scale controlled by $r_1$. This allows us
to conclude that for large $r_{\ast}$ we have
\beqs
\Lambda_0&=&\frac{M_Z^{\exp}}{\sqrt{2w_1}\varepsilon}\,,
\eeqs
and that in order to recover proper units we must multiply all the masses we find by this 
common scale.

Next, we compute 
\beqs
X_1(r)&=&e^{2A(r)}\int_0^r \di \omega e^{-4A(\omega)}X_0(\omega)^2\,,
\eeqs
and from it find that the precision parameter $\hat{S}$, which provided $r_{\ast}>0$ is given by 
\beqs
\hat{S}&=&\frac{\cos^2\theta_W \varepsilon^2}{4 w_1 \left(e^{2 r_{\ast}
   w_1} w_1-e^{2 (r_{\ast}+w_2)}+e^{2 (w_1
   r_{\ast}+r_{\ast}+w_2)}\right)^2}
\left(
e^{4 r_{\ast} w_1} (4 r_{\ast}+3) w_1^3-4 e^{2
   (w_1 r_{\ast}
   +r_{\ast}+w_2)} (2 r_{\ast}+1)
   w_1^2\right. \\ && \left.+4 e^{2 (2 w_1 r_{\ast}+r_{\ast}+w_2)} (2
   r_{\ast}+1) w_1^2+3 e^{4 (w_1
   r_{\ast}+r_{\ast}+w_2)}-4 e^{2 r_{\ast} (w_1+2)+4
   w_2} (2 r_{\ast} w_1+1)+e^{4 (r_{\ast}+w_2)}
   (4 r_{\ast} w_1+1)\right)\nonumber\,,
\eeqs
which for large values of $r_{\ast}$ becomes
\beqs
\hat{S}&\simeq&\frac{3\varepsilon^2\cos^2\theta_W}{4w_1}\,.
\eeqs
On the other hand, for $r_{\ast}\ll 0$ one recovers the pure AdS case
with unit curvature, which yields
\beqs
\hat{S}&\simeq & \frac{3\varepsilon^2\cos^2\theta_W}{4}\,.
\eeqs

At this point, we observe that for any reasonable values of $w_{1,2}$ and $r_{\ast}$,
this differs very little from the pure-AdS case.
Only when one chooses $\Phi_I\gg 1$ is the result suppressed by ${\cal O}(1)$ factors in
respect to the pure-AdS case, but then many of the current approximations cannot be trusted.

This means that one has to choose small values of $\varepsilon^2$ in order for 
the bound $\hat{S}<0.003$ to be satisfied, which in turns means that the 
parameters $W$ and $Y$ are automatically within their limit, being proportional to $\varepsilon^4$.
For example, if we choose $\varepsilon=0.07$, $\Phi_I=1$, $\Delta=1$, and a somewhat 
largish value of $r_{\ast}$
we obtain $\hat{S}\simeq 0.0026$, and $\Lambda_0\simeq 873$ GeV.

\section{Spectrum of heavy spin-1 states.}

We want now to compute the spectrum of the heavy resonances in the model.
In order to do so, we cannot rely on the formalism of the previous section, 
by making use of an expansion in $q^2$, but instead we need to 
reconsider the bulk equation for $\gamma(q,r)$.
We must compute the zeros of the complete $\pi_{V,A}$:
\beqs
\pi_V(q^2)&=&\lim_{r_2\rightarrow+\infty} -\varepsilon\left(r_2q^2-\frac{1}{\varepsilon^2}q^2+e^{2A(r_2)}\gamma(q,r_2)\right)\,,\\
\pi_A(q^2)&=&\lim_{r_2\rightarrow+\infty} -\varepsilon\left(r_2q^2-\frac{1}{\varepsilon^2}q^2+e^{2A(r_2)}\frac{1}{X(q,r_2)}\right)\,.
\eeqs

We do so approximately for $r_{\ast}\ll 0$, by assuming that $\Delta \Phi_I \leq 1$, in which case 
we can use the AdS result which yields:
\beqs
\pi_V&=&q^2+\varepsilon^2 q^2\left(\frac{\pi Y_0(q)}{2J_0(q)}-\gamma_E-\log\frac{q}{2}\right)\,,\\
\pi_A&=&q^2+\varepsilon^2 q^2\left(\frac{\pi Y_1(q)}{2J_1(q)}-\gamma_E-\log\frac{q}{2}\right)\,,
\eeqs
where $Y$ and $J$ are Bessel functions, and $\gamma_E$ is the Mascheroni constant.
The opposite case in which $r_{\ast}$ is very large, can be obtained by rescaling by a factor of $w_1$
all the dimensionful quantities,
so that for parametrically large $r_{\ast}$
\beqs
\pi_V&=&\frac{q^2}{w_1^2}+\varepsilon^2 \frac{q^2}{w_1^2}\left(\frac{\pi Y_0(q/w_1)}{2J_0(q/w_1)}-\gamma_E-\log\frac{q/w_1}{2}\right)\,,\\
\pi_A&=&\frac{q^2}{w_1^2}+\varepsilon^2 \frac{q^2}{w_1^2}\left(\frac{\pi Y_1(q/w_1)}{2J_1(q/w_1)}-\gamma_E-\log\frac{q/w_1}{2}\right)\,.
\eeqs

Under the assumption that $r_{\ast}$ be large,
the resulting spectrum consists of two towers of heavy states:
\beqs
M_{Vn}&\simeq &\left(n-\frac{1}{4}\right)\pi  w_1\,,\\
M_{An}&\simeq &\left(n+\frac{1}{4}\right)\pi w_1\,,
\eeqs
for $n=1,2\cdots$, plus a massless vectorial state $M_{V0}=0$ (the photon) and 
a light axial vector state $M_{a0}=M_Z\simeq \sqrt{2}\varepsilon w_1$.

Notice that the $w_1$-dependence does not drop even normalizing with the mass of
the $Z$ boson (as also clear from the fact that $\hat{S}$ does depend on $w_1$ and in general on $r_{\ast}$). 
The first excited state of the vectorial mode ($\r$ meson) has a mass
$M_{V1}\gsim 2.3$ TeV.  For the first excited axial-vector mode we have $M_{a_1}\gsim 3.6$ TeV.
Notice that one could lower these masses by allowing for larger values of $\Phi_I$,
but in this case some of our approximations might fail. We are not going to do so in the following.

\section{Scalar Spectrum.}

We can also compute the spectrum of scalar perturbations of the sigma-model.
We do so by using the formalism in~\cite{EP}, which requires to solve the equation
\SP{\label{Eq:diffeqWN}
	\Bigg[ e^{-4A} \left(  \partial_r + N \right) e^{4A}\left( \partial_r - N \right) 
	+ e^{-2A} q^2\Bigg] \mathfrak{a} = 0\,,
}
subject to the boundary conditions
\SP{\label{Eq:BCWinfty}
	&\left[ \frac{e^{2A}}{q^2} \frac{(\partial_{\Phi}W)^2}{W}\right] \left( \partial_r - N\right)\mathfrak a \Big|_{r_i} 
	=    \mathfrak a \Big|_{r_i}\,.
}
Notice that this comparatively simple results hold only because we took $m_i^2\rightarrow +\infty$.

For the heavy states, and as long as the departure from pure AdS is small, we can approximately solve
these equations by noticing that $N\simeq -\Delta$ is this limit, so that the heavy mass eigenstates can be found by solving
\beqs
\left((\partial_r+4-\Delta)(\partial_r+\Delta) +e^{-2r}q^2\right)\mathfrak{a}&=&0\,,\\
\left.\frac{}{}\mathfrak{a}\right|_{r_i}&=&0\,.
\eeqs
The general solution is
\beqs
\mathfrak{a}&=&e^{-2r}\left(b_1 J_{\Delta-2}(e^{-r}q)+b_2 Y_{\Delta-2}(e^{-r}q)\right)\,,
\eeqs
and the spectrum (for $r_1=0$ and $r_2\rightarrow+\infty$)
 is approximately given by the zeros of $J_{\Delta-2}(q)$, which for $\Delta=1$ are
 \beqs
 M_n&\simeq&\left(n+\frac{1}{4}\right)\pi\,.
 \eeqs
 
 This is the same result as for the axial-vector case.
 Notice instead that for very large values of $r_{\ast}$ the result is very different: 
 the spectrum is controlled by the effective scaling dimension computed at the IR fixed point, 
 which is very large, and hence the whole spectrum is shifted upwards.
 Furthermore, in this case one finds that that there is an overall small rescaling, due to $w_1$,
 in analogy to what happens for the heavy vectors.

More interesting is the case of the lightest state in the 
spectrum, which has to be treated more carefully.
As known from the literature, for $\Delta>2$, because we are going to take $r_2\rightarrow +\infty$,
this state would be exactly massless. More interesting is the case where $\Delta<2$,
which is what we focus on.
For extreme values of $r_{\ast}\rightarrow \pm\infty$, the dilaton becomes massless.
The complete $r_{\ast}$-dependence has been studied in details in~\cite{EP} and 
agrees with the PW case~\cite{KMPP}. It has a maximum, for $r_{\ast}$ close to zero, and falls-off
exponentially elsewhere. Because of this, it is actually possible in principle to use the 
mass of the light dilaton in order to extract information about the physical
scale $\Lambda_{\ast}\equiv\Lambda_0e^{r_{\ast}}$ that characterizes the background dynamics.

\begin{figure}[h]
\begin{center}
\begin{picture}(480,140)
\put(17,3){\includegraphics[height=4.5cm]{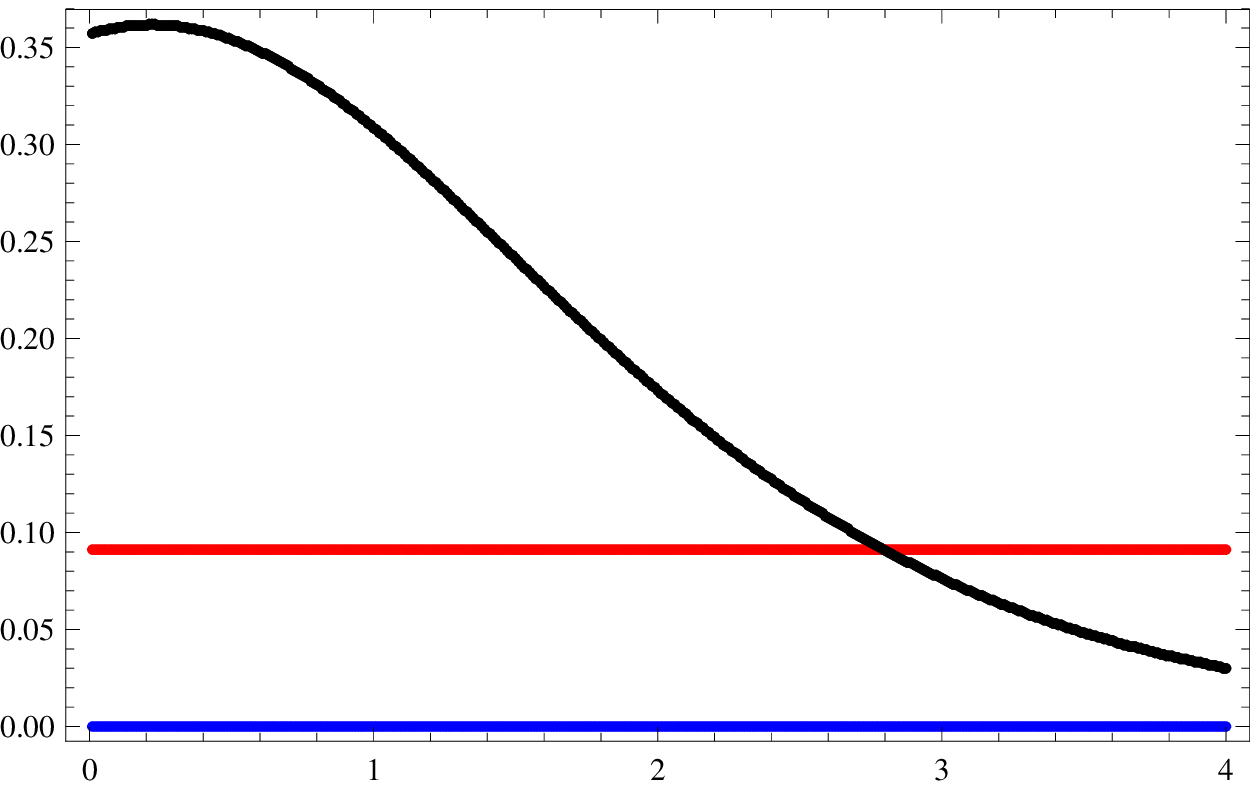}}
\put(236,3){\includegraphics[height=4.5cm]{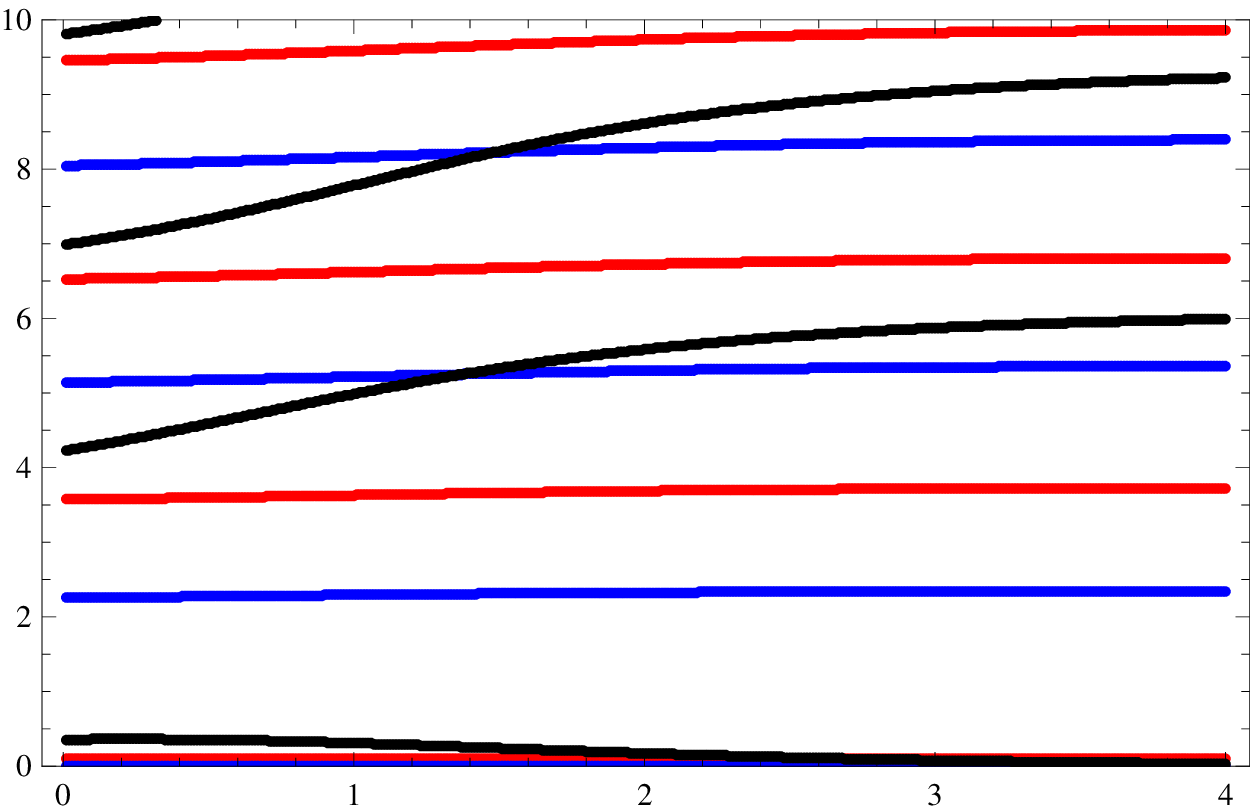}}
\put(205,0){$r_{\ast}$}
\put(405,0){$r_{\ast}$}
\put(5,100){$M$}
\put(228,100){$M$}
\end{picture} 
\caption{Left panel: the mass  (in TeV ) of the lightest states as a function of $r_{\ast}$,
for the choices $r_1=0$, $\Delta=1$ and $\Phi_I=1$. Right panel: the mass spectrum
 of vectorial, axial and scalar  states, for the same values of the parameters. 
 In blue are the vector resonances, in red the axial-vector states, and in black the scalars.}
\label{Fig:scalar}
\end{center}
\end{figure}

Let us conclude by numerically studying a few specific points
 in the  parameter space of this model,
as a way to  both test our approximations, and to illustrate what is going on. 
We fix $\Delta=1=e^{r_1}=\Phi_I$
and $\varepsilon=.07$ on the basis of the approximations we made
so far, and numerically compute all the relevant physical quantities.
The results are summarized in Table~\ref{Table:numbers} 
and in Fig.~\ref{Fig:scalar}.
The motivation for these choices is not particularly strong: there is no reason why $\Phi_I$ should be small,
and hence we take it to be of the same order as the confinement scale $e^{r_1}=1$. 
For the dimensionality, we choose it to be $\Delta=1$. For example
 this is what happens 
 in the Pilch-Warner flow~\cite{PW}, dual to the flow from ${\cal N}=4$ SYM to the Leigh-Strassler 
conformal fixed point~\cite{LS} (because of supersymmetry, in that case there are actually 
two active scalars in the five-dimensional theory, 
and their VEVs correspond to the deformation of the field theory by the presence of a mass terms for 
one of the fermions, with $\Delta=1$, and one of the scalars, with $\Delta=2$).
Notice how the very simple model studied here
 produces a light scalar spectrum which is qualitatively very
similar to what happens in the Pilch-Warner case~\cite{KMPP}, meaning that the
bulk geometry we choose is realistic.

\begin{table}
\begin{center}
\begin{tabular}{|c|c|c|c|c|c|c|c|}
\hline\hline
&\multicolumn{7}{c}{$\varepsilon=0.07$, $\Phi_I=1$, $\Delta=1$}\cr
\hline\hline
 $r_{\ast}$&1.5 & 2. & 2.2 & 2.3 & 2.4 & 2.5 & 2.6 \\
$M_Z$ & 0.0912 & 0.0912 & 0.0912 & 0.0912 & 0.0912 & 0.0912 & 0.0912 \\
$m$  &0.2410 & 0.1731 & 0.1490 & 0.1378 & 0.1272 & 0.1172 & 0.1078 \\
$\hat{S}$  &0.0026 & 0.0026 & 0.0026 & 0.0026 & 0.0026 & 0.0026 & 0.0026 \\
$\Lambda_{\ast}$& 3.9 & 6.5 & 7.9 & 8.7 & 9.6 & 10.7 & 11.8 \\
$M_{V1}$  &2.3 & 2.3 & 2.3 & 2.3 & 2.3 & 2.3 & 2.3 \\
$M_{V2}$  &5.3 & 5.3 & 5.3 & 5.3 & 5.3 & 5.3 & 5.3 \\
$M_{V3}$  &8.2 & 8.3 & 8.3 & 8.3 & 8.3 & 8.3 & 8.3 \\
$M_{A1}$  &3.7 & 3.7 & 3.7 & 3.7 & 3.7 & 3.7 & 3.7 \\
$M_{A2}$  &6.7 & 6.7 & 6.7 & 6.7 & 6.8 & 6.8 & 6.8 \\
 $M_{A3}$ &9.7 & 9.7 & 9.8 & 9.8 & 9.8 & 9.8 & 9.8 \\
$M_1$     &5.3 & 5.6 & 5.7 & 5.7 & 5.7 & 5.8 & 5.8 \cr
$M_{2}$   &8.2 & 8.6 & 8.7 & 8.8 & 8.8 & 8.9 & 8.9 \cr
\hline\hline
\end{tabular}
\end{center}
\caption{Numerical results for $\Delta=1=\Phi_I=e^{r_1}$, $\varepsilon=0.07$,
$r_2\rightarrow +\infty$ by varying $r_{\ast}$. All masses are in TeV.}
\label{Table:numbers}
\end{table}


Some brief comments.
Because $\Delta<2$, the mass of the light scalar state converges when 
$r_2\rightarrow +\infty$ and stays finite~\cite{EP}.
However, it is suppressed as by  $r_{\ast}$.
The comparison we make with the current bounds on the Higgs mass has to be taken with a grain of salt.
 Our main point is that  it is not particularly difficult to find regions of parameter space where the
dilaton has a mass in the allowed low-mass range $114<m<141$ GeV~\cite{LHC},
by dialing $r_{\ast}\sim 2.4$. 
It is also remarkable that there is an actual upper bound on $m\lsim 350$ GeV.
Because of the choice of $\varepsilon$, and hence because of the bound on $\hat{S}$,
there is a little desert between $m$ and the mass of the lightest $\r$ meson (around $2.3$ TeV).
Above this, one finds a rather complicated spectrum.
As expected, the scalar resonances all show the effect of the scale $r_{\ast}$,
while the spin-1 resonances are virtually insensitive to $r_{\ast}$.

\section{Critical discussion.}

Before concluding the paper, we want to alert the reader about two more intrinsic limitations of the
procedure we followed, and suggest to use caution in interpreting the results.

The model we propose contains a light scalar in the spectrum.
For illustration purposes, we compared its value, for a few choices of parameters, 
with the bounds on the mass of the 
standard-model Higgs particle. The reason why this is not completely 
unreasonable is that  we know that the light state is a (pseudo-)dilaton, 
and hence its couplings are going to be very similar to those of the 
Higgs particle (which is a dilaton itself). 
Namely, the light scalar couples to the light SM particles with couplings proportional to their masses 
(for fermions and gauge bosons), or  to the beta-functions computed by using the fields 
 that are heavier than the dilaton itself (for the gluon-gluon and photon-photon interactions).

There is no reason to expect the overall magnitude of such couplings to be precisely  the SM
one: in general the physical particle results from a mixing between the fluctuation of the 
metric and of the background scalar field, and this will result in a suppression of the couplings
(due to some mixing angle).
However, the new strongly coupled sector contains (techni-)fermions that are electrically charged.
This will enhance the coupling (and decay) to two photons.
Similar arguments hold for the production mechanism, which results from competing suppression and
enhancement mechanisms.
Comparing this composite dilaton to the standard model Higgs has to be done with 
some caution, both from the theoretical
point of view (a more precise and complete study of such couplings in this specific model
would be necessary), and from the experimental point of view (a very precise measurement 
of the coupling would be needed).

The second problem arises from the fact that, because we assumed that
symmetry-breaking be due to the IR boundary conditions (as opposed to the 
VEV of a bulk scalar), the only way to evade the bounds is to take 
$\varepsilon^2$ to be small. But in practice this means that 
the putative dual field theory is probably not at large-$N$, 
since $\varepsilon\propto \sqrt{N}$ (via unknown coefficients).
In turn, this means that the classical gravity approximation we are using is probably 
at the border of its intrinsic validity.
From the experimental perspective, this also means that the most accessible new state,
the techni-$\r$ meson, decays mostly into a final state with two $W$ bosons,
and that this decay is not suppressed, so that the resonance is going to be somewhat broad.
These two facts together might make the experimental detection of the techni-$\r$ quite challenging.

The reader should be aware of the fact that
both these problems could be addressed by refining the model,
and performing a more detailed study of all the couplings, 
which is beyond the purposes of this paper.

\section{Conclusions.}

The main physics message we want to convey, and that our paper exemplifies,
is remarkably simple. Neither the discovery nor the exclusion of a light scalar (Higgs)
particle provide, by themselves,  sufficient evidence allowing to disentangle weakly-coupled and
strongly-coupled origin of electro-weak symmetry breaking.

In the case where the Higgs particle is not found, it is not difficult to build models in which the couplings of such a state are suppressed, by assuming that several light scalars are present and that they mix non-trivially (an extreme example is for instance shown in~\cite{OGR}).
On the other hand, the example in this paper illustrates how nowadays
it is not difficult,  thanks to gauge/gravity dualities, to build models in which
a light dilaton (the couplings of which  resemble the ones of the Higgs particle)
is present in the spectrum, but EWSB is triggered by a strongly-coupled interaction.

In order to distinguish the two scenarios, a thorough exploration of the mass range up to several TeV 
is necessary, and the LHC program has the potential to do so.

Note added: while in the process of finalizing this paper, the LHC collaborations announced an update
on the results of the searches for the SM Higgs particle~\cite{new}. The results somewhat restrict the low-mass
window allowed by the data, where a low-significance hint of a positive signal is present.
These results leave our conclusions unchanged.

\vspace{1.0cm}
\section*{Acknowledgments}
The work of MP is supported in part  by WIMCS and by the STFC grant ST/G000506/1.
We would like to thank T. Hollowood and R. Lawrance for useful comments.


\begin{thebibliography}{99}


\bibitem{TC}
 S.~Weinberg,
  Phys.\ Rev.\ D {\bf 19}, 1277 (1979);
L.~Susskind,
  Phys.\ Rev.\ D {\bf 20}, 2619 (1979);
 S.~Weinberg,
  Phys.\ Rev.\ D {\bf 13}, 974 (1976).

\bibitem{WTC}
 B.~Holdom,
  Phys.\ Lett.\ B {\bf 150}, 301 (1985);
    K.~Yamawaki {\it et al.}
  Phys.\ Rev.\ Lett.\  {\bf 56}, 1335 (1986);
T.~W.~Appelquist {\it et al.}
  Phys.\ Rev.\ Lett.\  {\bf 57}, 957 (1986).


\bibitem{ETC}
 S.~Dimopoulos and L.~Susskind,
  Nucl.\ Phys.\ B {\bf 155}, 237 (1979);
 E.~Eichten and K.~D.~Lane,
  Phys.\ Lett.\ B {\bf 90}, 125 (1980).

\bibitem{reviews}
 R.~S.~Chivukula,
  arXiv:hep-ph/0011264;
K.~Lane,
  arXiv:hep-ph/0202255,
 C.~T.~Hill and E.~H.~Simmons,
  Phys.\ Rept.\  {\bf 381}, 235 (2003)
  [Erratum-ibid.\  {\bf 390}, 553 (2004)]
  [arXiv:hep-ph/0203079];
  A.~Martin,
  arXiv:0812.1841 [hep-ph];
   F.~Sannino,
  arXiv:0911.0931 [hep-ph];
  M.~Piai,
  arXiv:1004.0176 [hep-ph].

 \bibitem{dilaton}
  W.~A.~Bardeen {\it et al.}
  Phys.\ Rev.\ Lett.\  {\bf 56}, 1230 (1986);
    M.~Bando {\it et al.}
  Phys.\ Lett.\  B {\bf 178}, 308 (1986);
  Phys.\ Rev.\ Lett.\  {\bf 56}, 1335 (1986);
    B.~Holdom and J.~Terning,
  Phys.\ Lett.\  B {\bf 187}, 357 (1987);
  Phys.\ Lett.\  B {\bf 200}, 338 (1988).
  
\bibitem{dilatonpheno}
W.~D.~Goldberger {\it et al.}
  Phys.\ Rev.\ Lett.\  {\bf 100}, 111802 (2008);
  and 
  L.~Vecchi,
  arXiv:1002.1721 [hep-ph].

\bibitem{dilaton2}
D.~D.~Dietrich, F.~Sannino and K.~Tuominen,
  Phys.\ Rev.\  D {\bf 72}, 055001 (2005)
  [arXiv:hep-ph/0505059].
 T.~Appelquist and Y.~Bai,
  arXiv:1006.4375 [hep-ph].
  L.~Vecchi,
  arXiv:1007.4573 [hep-ph];
   K.~Haba, S.~Matsuzaki and K.~Yamawaki,
  Phys.\ Rev.\  D {\bf 82}, 055007 (2010)
  [arXiv:1006.2526 [hep-ph]];
        M.~Hashimoto and K.~Yamawaki,
  arXiv:1009.5482 [hep-ph].

\bibitem{dilaton5D}
 W.~D.~Goldberger and M.~B.~Wise,
  Phys.\ Lett.\  B {\bf 475}, 275 (2000)
  [arXiv:hep-ph/9911457];
  O.~DeWolfe, D.~Z.~Freedman, S.~S.~Gubser and A.~Karch,
  Phys.\ Rev.\  D {\bf 62}, 046008 (2000)
  [arXiv:hep-th/9909134].
C.~Csaki, M.~L.~Graesser and G.~D.~Kribs,
  Phys.\ Rev.\  D {\bf 63}, 065002 (2001)
  [arXiv:hep-th/0008151];
  L.~Kofman, J.~Martin and M.~Peloso,
  Phys.\ Rev.\  D {\bf 70}, 085015 (2004)
  [arXiv:hep-ph/0401189].
  
  \bibitem{Peskin}
 M.~E.~Peskin and T.~Takeuchi,
  Phys.\ Rev.\ D {\bf 46}, 381 (1992).

\bibitem{Barbieri}
R.~Barbieri, A.~Pomarol, R.~Rattazzi and A.~Strumia,
  Nucl.\ Phys.\ B {\bf 703}, 127 (2004)
  [arXiv:hep-ph/0405040].

  
\bibitem{AdSCFT}
  J.~M.~Maldacena,
  Adv.\ Theor.\ Math.\ Phys.\  {\bf 2}, 231 (1998)
  [Int.\ J.\ Theor.\ Phys.\  {\bf 38}, 1113 (1999)]
  [arXiv:hep-th/9711200];
  S.~S.~Gubser, I.~R.~Klebanov and A.~M.~Polyakov,
  Phys.\ Lett.\  B {\bf 428}, 105 (1998)
  [arXiv:hep-th/9802109];
  E.~Witten,
  Adv.\ Theor.\ Math.\ Phys.\  {\bf 2}, 253 (1998)
  [arXiv:hep-th/9802150].

\bibitem{Itzhaki:1998dd}
  N.~Itzhaki, J.~M.~Maldacena, J.~Sonnenschein and S.~Yankielowicz,
  Phys.\ Rev.\  D {\bf 58}, 046004 (1998)
  [arXiv:hep-th/9802042].

\bibitem{reviewAdSCFT}
 O.~Aharony, S.~S.~Gubser, J.~M.~Maldacena, H.~Ooguri and Y.~Oz,
  Phys.\ Rept.\  {\bf 323}, 183 (2000)
  [arXiv:hep-th/9905111].


 \bibitem{PS}
 J.~Polchinski and M.~J.~Strassler,
  arXiv:hep-th/0003136.
\bibitem{gppz}
  L.~Girardello, M.~Petrini, M.~Porrati and A.~Zaffaroni,
  Nucl.\ Phys.\  B {\bf 569}, 451 (2000)
  [arXiv:hep-th/9909047].
 
 
 
\bibitem{CO}
P.~Candelas, X.~C.~de la Ossa,
  Nucl.\ Phys.\  {\bf B342}, 246-268 (1990).
 
 
\bibitem{KW}
 I.~R.~Klebanov and E.~Witten,
  Nucl.\ Phys.\  B {\bf 536}, 199 (1998)
  [arXiv:hep-th/9807080].
 
  
  \bibitem{KT}
I.~R.~Klebanov and A.~A.~Tseytlin,
  Nucl.\ Phys.\  B {\bf 578}, 123 (2000)
  [arXiv:hep-th/0002159].

\bibitem{KS}
I.~R.~Klebanov and M.~J.~Strassler,
  JHEP {\bf 0008}, 052 (2000)
  [arXiv:hep-th/0007191];
M.~J.~Strassler,
  arXiv:hep-th/0505153.
 
 \bibitem{MN}
  J.~M.~Maldacena and C.~Nunez,
  Phys.\ Rev.\ Lett.\  {\bf 86}, 588 (2001).
  [arXiv:hep-th/0008001]. 
    See also
  A.~H.~Chamseddine and M.~S.~Volkov,
  Phys.\ Rev.\ Lett.\  {\bf 79}, 3343 (1997)
  [arXiv:hep-th/9707176].

 \bibitem{PT}
  G.~Papadopoulos and A.~A.~Tseytlin,
  Class.\ Quant.\ Grav.\  {\bf 18}, 1333 (2001)
  [arXiv:hep-th/0012034].


\bibitem{conifolds}
 A.~Butti, M.~Grana, R.~Minasian, M.~Petrini and A.~Zaffaroni,
  JHEP {\bf 0503}, 069 (2005)
  [arXiv:hep-th/0412187];
R.~Casero, C.~Nunez and A.~Paredes,
  Phys.\ Rev.\  D {\bf 73}, 086005 (2006)
  [arXiv:hep-th/0602027];
  Phys.\ Rev.\  D {\bf 77}, 046003 (2008)
  [arXiv:0709.3421 [hep-th]];
  C.~Hoyos-Badajoz, C.~Nunez and I.~Papadimitriou,
  Phys.\ Rev.\  D {\bf 78}, 086005 (2008)
  [arXiv:0807.3039 [hep-th]];
 C.~Nunez, I.~Papadimitriou and M.~Piai,
  arXiv:0812.3655 [hep-th].
    J.~Maldacena and D.~Martelli,
  JHEP {\bf 1001} (2010) 104
  [arXiv:0906.0591 [hep-th]].
   C.~Nunez, M.~Piai and A.~Rago,
  arXiv:0909.0748 [hep-th].
   R.~Minasian, M.~Petrini, A.~Zaffaroni,
  JHEP {\bf 1004}, 080 (2010).
  [arXiv:0907.5147 [hep-th]];
 N.~Halmagyi,
  [arXiv:1003.2121 [hep-th]];
  E.~Caceres, C.~Nunez, L.~A.~Pando-Zayas,
  [arXiv:1101.4123 [hep-th]];
J.~Gaillard, D.~Martelli, C.~Nunez and I.~Papadimitriou,
  Nucl.\ Phys.\  B {\bf 843}, 1 (2011)
  [arXiv:1004.4638 [hep-th]];
  D.~Elander, J.~Gaillard, C.~Nunez and M.~Piai,
  JHEP\ {\bf 1107}, 056  (2011)
  [arXiv:1104.3963 [hep-th]].
 E.~Conde, J.~Gaillard and A.~V.~Ramallo,
  JHEP\ {\bf 1110}, 023  (2011)
  [arXiv:1107.3803 [hep-th]].
  A.~Barranco, E.~Pallante and J.~G.~Russo,
  JHEP\ {\bf 1109}, 086  (2011)
  [arXiv:1107.4002 [hep-th]].
   L.~Anguelova,
  Nucl.\ Phys.\ B\ {\bf 843}, 429  (2011)
  [arXiv:1006.3570 [hep-th]].
   L.~Anguelova, P.~Suranyi and L.~C.~R.~Wijewardhana,
  Nucl.\ Phys.\ B\ {\bf 852}, 39  (2011)
  [arXiv:1105.4185 [hep-th]].

  \bibitem{consistentconifold}
  D.~Cassani and A.~F.~Faedo,
  arXiv:1008.0883 [hep-th];
I.~Bena, G.~Giecold, M.~Grana, N.~Halmagyi and F.~Orsi,
  arXiv:1008.0983 [hep-th].


\bibitem{SS}
  E.~Witten,
  Adv.\ Theor.\ Math.\ Phys.\  {\bf 2}, 505 (1998)
  [arXiv:hep-th/9803131],
  T.~Sakai and S.~Sugimoto,
  Prog.\ Theor.\ Phys.\  {\bf 113}, 843 (2005)
  [arXiv:hep-th/0412141].


\bibitem{AdSTC}
   D.~K.~Hong and H.~U.~Yee,
  Phys.\ Rev.\  D {\bf 74}, 015011 (2006)
  [arXiv:hep-ph/0602177];
 M.~Piai,
  arXiv:hep-ph/0608241,
  arXiv:hep-ph/0609104,
  arXiv:0704.2205 [hep-ph];
    K.~Haba, S.~Matsuzaki and K.~Yamawaki,
  arXiv:0804.3668 [hep-ph];
M.~Round,
  arXiv:1003.2933 [hep-ph].
J.~Hirn and V.~Sanz,
  Phys.\ Rev.\ Lett.\  {\bf 97}, 121803 (2006)
  [arXiv:hep-ph/0606086],
  JHEP {\bf 0703}, 100 (2007)
  [arXiv:hep-ph/0612239];
     C.~D.~Carone, J.~Erlich and J.~A.~Tan,
  arXiv:hep-ph/0612242;
M~Fabbrichesi, M.~Piai, L.~Vecchi
arXiv:0804.0124 [hep-ph];
  J.~Hirn, A.~Martin and V.~Sanz,
  arXiv:0807.2465 [hep-ph];
   M.~Round,
  Phys.\ Rev.\ D\ {\bf 84}, 013012  (2011)
  [arXiv:1104.4037 [hep-ph]].




\bibitem{PW}
  K.~Pilch and N.~P.~Warner,
  Adv.\ Theor.\ Math.\ Phys.\  {\bf 4}, 627 (2002)
  [arXiv:hep-th/0006066].
  
 
\bibitem{LS}
  R.~G.~Leigh and M.~J.~Strassler,
  Nucl.\ Phys.\  B {\bf 447}, 95 (1995)
  [arXiv:hep-th/9503121].

  
\bibitem{KMPP}
   S.~Prem Kumar, D.~Mateos, A.~Paredes and M.~Piai,
  JHEP\ {\bf 1105}, 008  (2011)
  [arXiv:1012.4678 [hep-th]].
  
  

  
  \bibitem{HR}
K.~Skenderis,
  Class.\ Quant.\ Grav.\  {\bf 19}, 5849 (2002)
  [arXiv:hep-th/0209067];
  I.~Papadimitriou and K.~Skenderis,
  arXiv:hep-th/0404176.

\bibitem{BHM}
 M.~Bianchi, M.~Prisco and W.~Mueck,
  JHEP {\bf 0311}, 052 (2003)
  [arXiv:hep-th/0310129];
  M.~Berg, M.~Haack and W.~Mueck,
  Nucl.\ Phys.\  B {\bf 736}, 82 (2006)
  [arXiv:hep-th/0507285];
    M.~Berg, M.~Haack and W.~Mueck,
  Nucl.\ Phys.\  B {\bf 789}, 1 (2008)
  [arXiv:hep-th/0612224].


\bibitem{EP}
 D.~Elander,
  JHEP {\bf 1003}, 114 (2010)
  [arXiv:0912.1600 [hep-th]];
D.~Elander and M.~Piai,
  arXiv:1010.1964 [hep-th].


\bibitem{ENP}
D.~Elander, C.~Nunez and M.~Piai,
  Phys.\ Lett.\  B {\bf 686}, 64 (2010)
  [arXiv:0908.2808 [hep-th]].

\bibitem{Round}
 M.~Round,
  Phys.\ Rev.\ D {\bf 82}, 053002 (2010)
  [arXiv:1003.2933 [hep-ph]].

\bibitem{LHC}
http://cdsweb.cern.ch/record/1399599/files/ATLAS-CONF-2011-157.pdf

\bibitem{OGR}
 O.~Bahat-Treidel, Y.~Grossman and Y.~Rozen,
  JHEP {\bf 0705}, 022 (2007)
  [hep-ph/0611162].
  
  \bibitem{new}
  ATLAS Note ATLAS-CONF-2011-163; CMS Physics Analysis Summary CMS PAS HIG-11-032.




  
 
  






\end{thebibliography}
\end{document}